\documentclass[prl,twocolumn,letterpaper,showpacs,amsmath]{revtex4}
\usepackage{graphicx}
\usepackage{color}

\begin{document}
\title{Cotunneling-mediated transport through excited states\\ in the Coulomb blockade regime}
\author{R.~Schleser}
\author{T.~Ihn}
\author{E.~Ruh}
\author{K.~Ensslin}
\affiliation{Solid State Physics Laboratory, ETH Z\"urich, 8093 Z\"urich, Switzerland}
\author{M.~Tews}
\author{D.~Pfannkuche}
\affiliation{I.~Institut f\"ur Theoretische Physik, Universit\"at Hamburg, Jungiusstrasse 9, D-20355 Hamburg, Germany}
\author{D.~C.~Driscoll}
\author{A.~C.~Gossard}
\affiliation{Materials Department, University of California, Santa Barbara California 93106}
\date{\today}
\begin{abstract}
We present finite bias transport measurements on a few-electron quantum dot. In the Coulomb blockade regime, strong signatures of inelastic cotunneling occur which can directly be assigned to excited states observed in the non-blockaded regime. In addition, we observe structures related to sequential tunneling through the dot, occuring after it has been excited by an inelastic cotunneling process.
We  explain our findings using transport calculations within the real-time Green's function approach,
including diagrams up to fourth order in the tunneling matrix elements.
\end{abstract}
\maketitle

\newcommand{\nf}{\normalfont}
\newcommand{\cm}{\,\text{cm}}
\newcommand{\nm}{\,\text{nm}}
\newcommand{\muV}{\,\mu\text{V}}
\newcommand{\mV}{\,\text{mV}}
\newcommand{\V}{\,\text{V}}
\newcommand{\fA}{\,\text{fA}}
\newcommand{\meV}{\,\text{meV}}
\newcommand{\mueV}{\,\mu\text{eV}}
\newcommand{\eVV}{\,\text{eV}/\text{V}}
\newcommand{\Hz}{\,\text{Hz}}
\newcommand{\xHz}{\text{Hz}}
\newcommand{\K}{\,\text{K}}
\newcommand{\mK}{\,\text{mK}}
\newcommand{\T}{\,\text{T}}
\newcommand{\kB}{k_\text{B}}

\renewcommand{\topfraction}{0.9}
\renewcommand{\bottomfraction}{0.9}
\renewcommand{\dbltopfraction}{0.9}
\setcounter{totalnumber}{5}
\setcounter{topnumber}{3}
\setcounter{bottomnumber}{3}
\setcounter{dbltopnumber}{3}


In a quantum dot in the Coulomb blockade regime, the energy gap related to
the charging energy becomes larger than $k_{\rm B}T$ and
sequential tunneling transport involving only dot ground states is exponentially suppressed (see e.g.~\cite{van-Houten1992}).
Transport is dominated by cotunneling \cite{Averin1990}.
Elastic cotunneling, prevalent at low bias voltages,
involves virtual tunneling of one electron through the dot via a higher-energy state
and leaves the dot in the ground state.
Inelastic processes imply correlated tunneling of two electrons, leaving the dot in an excited state
with energy $\Delta$ above the ground state. Inelastic cotunneling sets in once the bias energy
$eV_\text{bias}\geq\Delta$.
Recently, Golovach and Loss have presented a theoretical analysis of the interplay
between cotunneling and sequential tunneling in a double dot system \cite{Golovach2004}.
Experimental investigations involving cotunneling have
been performed on metallic \cite{Geerligs1990, Hanna1992, Eiles1992}
and semiconducting \cite{Glattli1991, Pasquier1993, Cronenwett1997} systems containing a
large number of electrons.
Signatures of inelastic cotunneling
in a transport measurement have been observed in investigations on
small vertical semiconductor quantum dots \cite{De-Franceschi2001,Yamada2003},
and in single-walled \cite{Nygard2000} and multiwalled \cite{Buitelaar2002} carbon
nanotubes, all containing well separated energy levels.

In the following, we first present finite bias transport measurements through
a quantum dot, showing structure
outside as well as within the Coulomb blockade regime.
In the second part, we present a theoretical analysis of our results
based on transport calculations using the real-time Green's function approach.

\begin{figure}[t!b]
\includegraphics[width=3.075in]{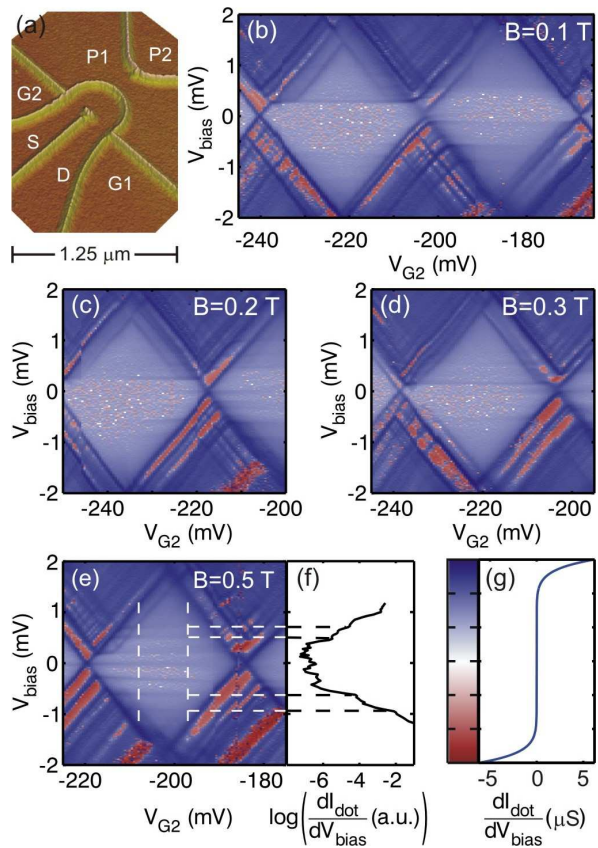}
\caption{(Color online)
(a) AFM micrograph of structure with designations of 
gates: source (S) and drain (D) of the quantum dot; lateral gates 
G1 and G2 to control the coupling of the dots to the reservoirs; 
Plunger gates P1 and P2 to tune the number of electrons on the dot.
(b) Finite bias
measurement of the dot's differential conductance $dI/dV$ at $B=0.1\T$.
(c)-(e) Finite bias transport measurement at $B=0.2\T$, $B=0.3\T$, and $B=0.5\T$, respectively.
Note the different vertical distances between the
diamond edge and the diagonal structures inside the Coulomb blockaded region.
(f) Single trace of the bias dependence of the current in the blockaded regime,
obtained by averaging over the gate voltage range marked in (f) by two vertical dashed lines.
(g) Color scale used in (b)-(e): To show all features inside and outside
the Coulomb blockaded region, our data is presented in a highly nonlinear color scale. The
plotted  quantity is calculated as $\sqrt[10]{\left|x\right|}\cdot{\rm sign}(x)$ with $x=dI/dV\cdot 1\,\Omega$.
\label{fig1}}
\end{figure}

The sample (see Fig.~\ref{fig1}(a)) was fabricated 
by surface probe lithography \cite{Held1999, Luscher1999}
on a GaAs/Al$_{0.3}$Ga$_{0.7}$As heterostructure, 
containing a two-dimensional electron gas (2DEG) $34\nm$ below the 
surface as well as a backgate (BG) $1400\nm$ below the 2DEG.
The unstructured 2DEG had a mobility of
$(3.5\pm 0.5)\cdot 10^{5}\cm^{2}/{\rm Vs}$ and a density of $(4.6\pm 0.5)\cdot 10^{11}\cm^{-2}$
at a temperature $T=4.2\K$ and a BG voltage $V_{\rm BG}=-0.5\V$.

All measurements were performed in a dilution refrigerator
with a base temperature of $80 \mK$.
Negative voltages were applied to the surrounding gates
(see Fig.~\ref{fig1}(a)) and to the back gate, to tune the charge 
on the dot and the transparency of its tunnel barriers.
The bias voltage $V_\text{\textnormal bias}$ was 
applied symmetrically (with respect to ground)
across the dot between source (S) and drain (D).
The dc transport current was measured and numerically differentiated.
An estimated charging energy $E_{\rm c}\approx 1.6\meV$
and a single level spacing $\Delta\approx 0.3\meV$ were extracted.

Figure \ref{fig1} shows measurements of finite bias 
differential conductance $dI/dV$ on a strongly nonlinear scale.
Figs.~\ref{fig1}(b)-(e) contain measurements of differential conductance vs.~both gate voltage and bias voltage.
Measurements were performed at different magnetic fields in order to vary
the wave functions inside the quantum dot and their coupling to the reservoirs.
Inside the diamond-shaped regions, i.e. in the Coulomb blockaded regime,
we observe horizontal (constant bias) structures.
At the diamond boundary, the horizontal lines seamlessly
join some of the most prominent diagonal lines in the non-blockaded region.
In Figure \ref{fig1}(f), an averaged trace of the current vs.~bias
voltage is presented, showing the position of these kinks more precisely.

For positive bias, e.g.~in Fig.~\ref{fig1}(b), additional structure
inside the diamond is observed:
for bias voltages above the well-resolved horizontal threshold line, diagonal
lines parallel to the diamond edges appear. In our measurements,
this feature remains visible at all magnetic fields measured
(from $B=0.1\T$ to $B=0.5\T$ in steps of $0.1\T$, see Figs.~\ref{fig1}(c)-(e) for more examples).
The vertical distance between the diagonal lines and the diamond edge
is identical for positively (left of diamond) and negatively sloped lines.
When extended towards higher or towards negative voltages, most of the diagonal
lines apparently join prominent lines in the non-blockaded regime.

A closer examination of the structures
reveals a connection between two energy scales visible inside the blockaded region (see Figure~\ref{fig3}(c) for an illustration):
an extension of each diagonal line intersects the zero bias line at a certain point (point A in Fig.~\ref{fig3}(c)).
Connecting this point to the diamond edge at its intersection with the closest horizontal line
(point B in Fig.~\ref{fig3}(c))
yields an extension of a diagonal line (of opposite slope) in the non-blockaded regime.
This remains valid at different magnetic fields, where
the vertical distance between the diagonal line and the diamond edge varies by a factor of about 2.

While we observe the horizontal structures for at least 12 consecutive Coulomb blockade diamonds,
the diagonal features are only seen for a maximum of two neighbouring diamonds up to now.
On reducing the transparencies of the dot's tunnel barriers, the amplitude
of the cotunneling currents become comparable to the minimum current resolution,
and the structures inside the Coulomb diamonds gradually disappear.

We interpret our findings as follows: the horizontal lines in the blockaded regime mark
the onset of inelastic cotunneling connected to specific excited states. The distance from the
zero-bias line corresponds to the single-particle level spacing of these states with respect to the
ground state.
At the intersection points at the border of the Coulomb diamonds, a direct mapping can be made 
of the excited states that contribute measurably to inelastic cotunneling and those that open additional
transport channels in the non-blockaded, finite-bias regime.
The horizontal line close to zero bias in Fig.~\ref{fig1}(e) suggests the presence of a state with low
excitation energy contributing to inelastic cotunneling.

The most unconventional features observed are the diagonal
lines inside the Coulomb blockaded regions. The fact that they have
the same slope as the diamond edges suggests that they are connected to the alignment of an energy level
with source (negative slope) or drain (positive slope).

To verify this hypothesis, we have performed transport calculations within the
real-time Green's function approach \cite{Schoeller94:18436} including all
cotunneling diagrams, i.e.~all diagrams to fourth order in the tunneling
matrix elements \cite{Tews04:xx}. In the following, the results for a quantum dot
with a simple level structure are discussed.

The simplest quantum dot which shows signatures of its excitation spectrum in
the Coulomb blockade regime is described by an Anderson Hamiltonian
$\hat{H}_{\rm D}$ having two non-degenerate single-particle levels $E_1$ and $E_2$ and the excitation energy $\Delta=E_2-E_1$:
\begin{eqnarray}
  \hat{H} &=& \hat{H}_\text{D} +\hat{H}_\text{R} +\hat{H}_\text{T} \\
  \hat{H}_\text{D} &=& \sum_{l=1,2} E_l c_l^+c_l^{} + U c_1^+c_1^{}c_2^+c_2^{}
            + e\alpha_\text{G} V_\text{G} \hat{N}_\text{D} \label{eqn_H_D}
            \label{eqn_H_T}
\end{eqnarray}
The annihilation (creation) operator $c_l^{(+)}$ annihilates (creates) an
electron of state $l$ in the quantum dot. Coulomb interaction is described by
the second term of (\ref{eqn_H_D}) leading to an additional interaction energy
$U$ whenever the dot is occupied by two electrons. The four possible states of
the isolated quantum dot are labeled as follows:
$|0,0\rangle$ denotes the empty dot, $|1,0\rangle$ the single-particle ground
state, $|1,1\rangle$ the excited single-particle state, and $|2,0\rangle$ the
two-particle state. Here we assume that the applied gate voltage $V_\text{G}$  leads
to a constant electrostatic potential described by the third term of the
quantum dot Hamiltonian. In this term, $\hat{N}_\text{D}=\sum_{l}c_{l}^+c_{l}$ is the
number operator for the dot electrons and $\alpha_\text{G}$ the electrostatic lever arm of the
gate electrode.
The coupling of the quantum dot to two reservoirs is described by the
reservoir ($\hat{H}_\text{R}$) and the tunneling Hamiltonian
($\hat{H}_\text{T}$). They are of the conventional form (see e.g. \cite{Tews04:xx}, \cite{cohen:04:1962})
with the reservoir electrons being treated
as non-interacting except
for an overall self-consistent potential \cite{jauho:1998}, owing to the high
density of states in source and drain contacts. 
For simplicity, we
assume for the following the absolute value of the complex tunneling matrix
elements to be independent of all quantum numbers with a complex phase which is random with
respect to the direction of the reservoir electrons wave vector. 

In order to calculate the non-equilibrium transport properties for finite
transport voltages $V_\text{SD}$, we use the real-time transport theory developed by
Schoeller~\emph{et~al.}~\cite{Schoeller94:18436}. Following the steps of
this theory one can trace out the reservoir degrees of freedom and derive a
formally correct equation of motion for the reduced density matrix of the
quantum dot system which under steady state conditions transforms into 
\begin{eqnarray}
  \frac{i}{\hbar}\left(E_s-E_s'\right)P^{\text st}_{ss'}=
  \sum_{s_1s_1'}  P^\text{st}_{s_1s_1'}
  \int_{-\infty}^{0}dt'\Sigma_{ss's_1s_1'}(0,t'). \label{eqn:statRDM}
\end{eqnarray}
Here $P_{ss'}$ denotes a matrix element of the reduced dot density matrix with
the (few)-particle states $|s\rangle$ and $|s'\rangle$ of the isolated quantum
dot. The kernel $\Sigma_{ss's_1s_1'}$ represents a generalized transition rate
involving the relevant tunneling processes.
Within the same formalism one can also calculate the tunneling current
expectation value for the steady state 
\begin{eqnarray}
  \langle I_r^\text{st}\rangle = -e \sum_{ss_1s_1'} P_{s_1s_1'}^\text{st}
  \int_{-\infty}^{0}dt'\Sigma^r_{sss_1s_1'}(0,t').
  \label{eqn:curr_st}
\end{eqnarray}
In the sequential tunneling approximation, a ``tunneling in'' process
is only possible if a reservoir electron matches the energy required to
charge the quantum dot by a further electron. Generally, this energy is
given by $E_s-E_{s'}$ for a transition between the
state $|s\rangle$ (where $|s\rangle=|N,j\rangle$ is the $j$th $N$-particle state)
and the $(N+1)$-particle state $|s'\rangle$ and is in the
following called transport channel and denoted by $\mu(s;s')$. For the quantum
dot described by (\ref{eqn_H_D}), four transport channels exist.
In the Coulomb blockade regime, where transport in lowest order is exponentially
supressed, the electrochemical potentials of both
reservoirs are in between the two transport channels associated with two
groundstates: $\mu(2,0;1,0) \geq \mu_r \geq \mu(1,0;0,0)$. Going one step
further and calculating the kernel of (\ref{eqn:statRDM}) 
in fourth order, already 64 qualitativly different terms occur, describing
so-called cotunneling processes in which two electrons participate coherently in
a tunneling process. In the following we study the resulting differential conductance
including consistently all cotunneling contributions
\footnote{An advantage of the real-time transport theory is that the kernel
$\Sigma_{ss's_2s_2'}(t,t')$ is well defined and can be expressed
systematically order by order in the tunnel coupling strength
$\frac{\Gamma}{k_{\rm B}T}=\frac{D|{\cal T}|^2}{h\hspace{1mm}k_{\rm B}T}$.}.

\begin{figure}[tb]      
  \includegraphics[width=3.075in]{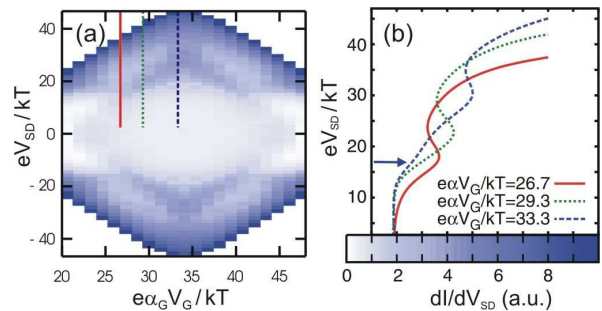}
  \caption
  {(Color online)
        (a) Calculated charging diagram within the Coulomb blockade
      regime including the cotunneling contribution. The peak
      conductances due to resonant sequential tunneling (also
      responsible for the increasing background in (b)) have been
      ``cut'' from the color scale (outer white regions). Solid, dotted
      and dashed lines correspond to traces in (b).
   (b) Cotunneling signatures within the Coulomb blockade
    regime. Shown is the differential conductance versus the applied
    transport voltage for three different gate voltages.
    The arrow marks the onset of inelastic cotunneling
    for the dashed trace.
    The parameters used for this calculation are: $\Gamma=0.1k_{\rm B}T$,
    $\delta=14.67k_{\rm B}T$, and $U=52k_{\rm B}T$.
    Bottom: color scale for (a).}
     \label{fig:CB_co_cd}
     \label{fig:CB_co}
     \label{fig2}
\end{figure}

\begin{figure}[tb]
      \includegraphics[width=3.075in]{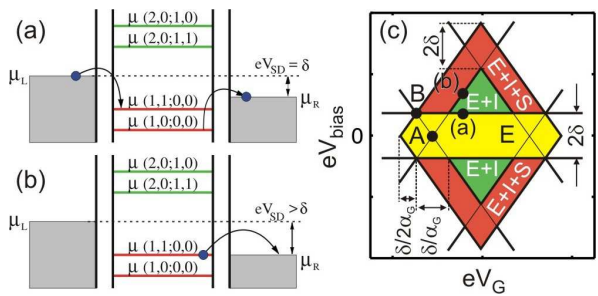}
      \caption
        {(Color online) (a/b) Relative position of transport channels and
        contact electrochemical potentials for different bias voltages.
      (c) Schematic picture of the various tunneling regimes
        within the Coulomb blockade.
        A and B label the marked intersection points. (a) and (b) mark the
        gate voltage values of the corresponding diagrams.
        E is for elastic and I for inelastic cotunneling,
        S for sequential tunneling.
      \label{fig:CB_co_schematic_cd}
      \label{fig:GoldenRuleRates_co}
      \label{fig3}}
  \end{figure}

In Fig.~\ref{fig:CB_co}(b) the differential conductance as a function of the
applied source-drain voltage $V_\text{SD}$ is shown
within the Coulomb blockade regime, i.e., the electro-chemical potentials for
$V_\text{SD}=0$ are energetically in between the two ground state
channels $\mu(2,0;1,0)$ and $\mu(1,0;0,0)$. 
For small voltages, all three traces start with the same value
and at least the traces for $e\alpha_\text{G} V_\text{G}/k_{\rm B}T=29.3$ and $e\alpha_\text{G} V_\text{G}/k_{\rm B}T=33.3$
stay constant for small transport voltages. This constant and finite
differential conductance can be attributed to elastic cotunneling by virtual
tunneling through either the vacuum or the two-particle state. Additionally,
for all three traces a peak is found which shifts linearly to higher source-drain voltages with increasing gate
voltage. In contrast to the traces at
lower gate voltages, the differential conductance of the highest gate voltage
($e\alpha_\text{G} V_\text{G}/k_{\rm B}T=33.3$) shows an additional step (see arrow in Fig.~\ref{fig:CB_co}(b))
emerging at the source-drain
voltage $eV_\text{SD}/k_{\rm B}T\approx 15$, which corresponds to the excitation energy
$\delta$. This step
is strongly smeared due to temperature and especially due
to the overlap with the peak occurring at higher voltages (also responsible
for the strong increase towards higher bias voltages).

  In Fig.~\ref{fig:GoldenRuleRates_co}(a), the
  relative position of the transport channels with respect to the
  electrochemical potentials in the reservoirs is shown at the
  parameters where the step occurs ($eV_\text{SD}=\delta$ and
  $e\alpha_\text{G} V_\text{G}/k_{\rm B}T=33.3$).  In this situation, inelastic cotunneling
  becomes possible in which the quantum dot becomes excited during the
  tunneling process.  This additional (to the elastic process)
  cotunneling process leads to a step in the differential conductance
  \cite{funabashi:01:1999} which corresponds to the experimentally observed
  steps showing as horizontal lines in Fig.~1. Eventually, by further
  increasing the 
  source-drain voltage $eV_\text{SD} > \delta$, the electrochemical
  potential of the drain reservoir becomes resonant with the transport
  channel $\mu(1,1;0,0)$ (sketched in Fig.~\ref{fig:GoldenRuleRates_co}(b)).
  Other than in the sequential tunneling approximation, where the
  $|1,1\rangle$ state cannot be occupied due to the Coulomb blockade
  effect, inelastic cotunneling allows to occupy this excited state
  and the resonant channel leads to a peak in the differential
  conductance. Due to the smaller cotunneling rate, the peak is
  lower as compared to the corresponding peak beyond the Coulomb
  blockade regime.
  
  For lower gate voltages, the peak moves to lower
  source-drain voltages and eventually merges with the step at
  $eV_\text{SD}=\delta$.
  For even lower
  gate voltages, the channel $\mu(1,1;0,0)$ is already within the
  transport window at the source-drain voltage $eV_\text{SD}=\delta$ needed
  to allow for the inelastic cotunneling process.

  Combining all these processes, various tunneling regimes within the
  Coulomb blockade can be identified as sketched in
  Fig.~\ref{fig:CB_co_schematic_cd}(c). For
  $eV_\text{SD}<\delta$, transport is dominated
  by elastic cotunneling, leading to a constant offset of the
  differential conductance. For gate-voltages in the vicinity of the
  Coulomb blockade center and $eV_\text{SD}>\delta$, a regime where elastic
  and inelastic cotunneling occur is found. In the remaining outer
  regime, sequential tunneling through the excited single-particle
  state is also possible. At the border of this regime, a peak occurs in the
  differential conductance. All described features are also found in
  the calculated charging diagram including cotunneling (shown in
  Fig.~\ref{fig:CB_co_cd}(a)).

While it is not entirely clear to us why the induced sequential tunneling
contributions have not been observed before, we can identify a few requirements:
First of all, the charging energy has to be large enough,
$E_\text{c}>2\delta$. This can be directly seen from Fig. \ref{fig3}(c), and
it explains why the effect has not been observed e.g. in \cite{De-Franceschi2001}.
In addition, a sufficient level spacing $\delta>\!>\kB T$ is required
so that the effect is not smeared out by temperature.
We speculate that the strongly asymmetric tunnel barrier configuration
used in the present experiment may also
enhance the visibility of these features.

Elastic cotunneling has earlier been identified as a possible source of uncertainty
in the operation of single-electron devices (see e.g. \cite{Averin1990}).
Our results show that the inelastic contributions can become more prominent,
especially if the induced sequential tunneling is taken into account.
It follows that in an application relying on Coulomb
blockade in quantum dots, e.g. in quantum information processing,
the bias must be kept small in comparison to the lowest
excitation energy.

We thank L. Kouwenhoven for valuable discussions.
Financial support from the Swiss National Science Foundation and the
NCCR {\lq\lq Nanoscale Science\rq\rq} is gratefully acknowledged.
M.T.~and D.P.~acknowledge financial support of the
Deutsche Forschungsgemeinschaft via SFB 508 \lq\lq Quantum Materials\rq\rq.

\bibliographystyle{apsrev}

\begin{thebibliography}{21}
\expandafter\ifx\csname natexlab\endcsname\relax\def\natexlab#1{#1}\fi
\expandafter\ifx\csname bibnamefont\endcsname\relax
  \def\bibnamefont#1{#1}\fi
\expandafter\ifx\csname bibfnamefont\endcsname\relax
  \def\bibfnamefont#1{#1}\fi
\expandafter\ifx\csname citenamefont\endcsname\relax
  \def\citenamefont#1{#1}\fi
\expandafter\ifx\csname url\endcsname\relax
  \def\url#1{\texttt{#1}}\fi
\expandafter\ifx\csname urlprefix\endcsname\relax\def\urlprefix{URL }\fi
\providecommand{\bibinfo}[2]{#2}
\providecommand{\eprint}[2][]{\url{#2}}

\bibitem[{\citenamefont{van Houten et~al.}(1992)\citenamefont{van Houten,
  Beenakker, and Staring}}]{van-Houten1992}
\bibinfo{author}{\bibfnamefont{H.}~\bibnamefont{van Houten}},
  \bibinfo{author}{\bibfnamefont{C.~W.~J.} \bibnamefont{Beenakker}},
  \bibnamefont{and} \bibinfo{author}{\bibfnamefont{A.~A.~M.}
  \bibnamefont{Staring}}, in \emph{\bibinfo{booktitle}{Single Charge Tunneling:
  Coulomb blockade Phenomena in Nanostructures}}, edited by
  \bibinfo{editor}{\bibfnamefont{H.}~\bibnamefont{Grabert}} \bibnamefont{and}
  \bibinfo{editor}{\bibfnamefont{M.~H.} \bibnamefont{Devoret}}
  (\bibinfo{publisher}{Plenum Press and NATO Scientific Affairs Division, New
  York}, \bibinfo{year}{1992}), vol. \bibinfo{volume}{294} of
  \emph{\bibinfo{series}{NATO ASI Series, Series B: Physics}}, p.
  \bibinfo{pages}{167}.

\bibitem[{\citenamefont{Averin and Nazarov}(1990)}]{Averin1990}
\bibinfo{author}{\bibfnamefont{D.~V.} \bibnamefont{Averin}} \bibnamefont{and}
  \bibinfo{author}{\bibfnamefont{Y.~V.} \bibnamefont{Nazarov}},
  \bibinfo{journal}{Phys. Rev. Lett.} \textbf{\bibinfo{volume}{65}},
  \bibinfo{pages}{2446} (\bibinfo{year}{1990}).

\bibitem[{\citenamefont{Golovach and Loss}(2004)}]{Golovach2004}
\bibinfo{author}{\bibfnamefont{V.~N.} \bibnamefont{Golovach}} \bibnamefont{and}
  \bibinfo{author}{\bibfnamefont{D.}~\bibnamefont{Loss}},
  \bibinfo{journal}{Phys. Rev. B} \textbf{\bibinfo{volume}{69}},
  \bibinfo{pages}{245327} (\bibinfo{year}{2004}).

\bibitem[{\citenamefont{Geerligs et~al.}(1990)\citenamefont{Geerligs, Averin,
  and Mooij}}]{Geerligs1990}
\bibinfo{author}{\bibfnamefont{L.~J.} \bibnamefont{Geerligs}},
  \bibinfo{author}{\bibfnamefont{D.~V.} \bibnamefont{Averin}},
  \bibnamefont{and} \bibinfo{author}{\bibfnamefont{J.~E.} \bibnamefont{Mooij}},
  \bibinfo{journal}{Phys. Rev. Lett.} \textbf{\bibinfo{volume}{65}},
  \bibinfo{pages}{3037} (\bibinfo{year}{1990}).

\bibitem[{\citenamefont{Hanna et~al.}(1992)\citenamefont{Hanna, Tuominen, and
  Tinkham}}]{Hanna1992}
\bibinfo{author}{\bibfnamefont{A.~E.} \bibnamefont{Hanna}},
  \bibinfo{author}{\bibfnamefont{M.~T.} \bibnamefont{Tuominen}},
  \bibnamefont{and} \bibinfo{author}{\bibfnamefont{M.}~\bibnamefont{Tinkham}},
  \bibinfo{journal}{Phys. Rev. Lett.} \textbf{\bibinfo{volume}{68}},
  \bibinfo{pages}{3228} (\bibinfo{year}{1992}).

\bibitem[{\citenamefont{Eiles et~al.}(1992)\citenamefont{Eiles, Zimmerli,
  Jensen, and Martinis}}]{Eiles1992}
\bibinfo{author}{\bibfnamefont{T.~M.} \bibnamefont{Eiles}},
  \bibinfo{author}{\bibfnamefont{G.}~\bibnamefont{Zimmerli}},
  \bibinfo{author}{\bibfnamefont{H.~D.} \bibnamefont{Jensen}},
  \bibnamefont{and} \bibinfo{author}{\bibfnamefont{J.~M.}
  \bibnamefont{Martinis}}, \bibinfo{journal}{Phys. Rev. Lett.}
  \textbf{\bibinfo{volume}{69}}, \bibinfo{pages}{148} (\bibinfo{year}{1992}).

\bibitem[{\citenamefont{Glattli et~al.}(1991)\citenamefont{Glattli, Pasquier,
  Meirav, Williams, Jin, and Etienne}}]{Glattli1991}
\bibinfo{author}{\bibfnamefont{D.~C.} \bibnamefont{Glattli}},
  \bibinfo{author}{\bibfnamefont{C.}~\bibnamefont{Pasquier}},
  \bibinfo{author}{\bibfnamefont{U.}~\bibnamefont{Meirav}},
  \bibinfo{author}{\bibfnamefont{F.~I.~B.} \bibnamefont{Williams}},
  \bibinfo{author}{\bibfnamefont{Y.}~\bibnamefont{Jin}}, \bibnamefont{and}
  \bibinfo{author}{\bibfnamefont{B.}~\bibnamefont{Etienne}},
  \bibinfo{journal}{Z. Phys. B} \textbf{\bibinfo{volume}{85}},
  \bibinfo{pages}{375} (\bibinfo{year}{1991}).

\bibitem[{\citenamefont{Pasquier et~al.}(1993)\citenamefont{Pasquier, Meirav,
  Williams, Glattli, Jin, and Etienne}}]{Pasquier1993}
\bibinfo{author}{\bibfnamefont{C.}~\bibnamefont{Pasquier}},
  \bibinfo{author}{\bibfnamefont{U.}~\bibnamefont{Meirav}},
  \bibinfo{author}{\bibfnamefont{F.~I.~B.} \bibnamefont{Williams}},
  \bibinfo{author}{\bibfnamefont{D.~C.} \bibnamefont{Glattli}},
  \bibinfo{author}{\bibfnamefont{Y.}~\bibnamefont{Jin}}, \bibnamefont{and}
  \bibinfo{author}{\bibfnamefont{B.}~\bibnamefont{Etienne}},
  \bibinfo{journal}{Phys. Rev. Lett.} \textbf{\bibinfo{volume}{70}},
  \bibinfo{pages}{69} (\bibinfo{year}{1993}).

\bibitem[{\citenamefont{Cronenwett et~al.}(1997)\citenamefont{Cronenwett,
  Patel, Marcus, Campman, and Gossard}}]{Cronenwett1997}
\bibinfo{author}{\bibfnamefont{S.~M.} \bibnamefont{Cronenwett}},
  \bibinfo{author}{\bibfnamefont{S.~R.} \bibnamefont{Patel}},
  \bibinfo{author}{\bibfnamefont{C.~M.} \bibnamefont{Marcus}},
  \bibinfo{author}{\bibfnamefont{K.}~\bibnamefont{Campman}}, \bibnamefont{and}
  \bibinfo{author}{\bibfnamefont{A.~C.} \bibnamefont{Gossard}},
  \bibinfo{journal}{Phys. Rev. Lett.} \textbf{\bibinfo{volume}{79}},
  \bibinfo{pages}{2312} (\bibinfo{year}{1997}).

\bibitem[{\citenamefont{De~Franceschi et~al.}(2001)\citenamefont{De~Franceschi,
  Sasaki, Elzerman, Van-Der-Wiel, Tarucha, and
  Kouwenhoven}}]{De-Franceschi2001}
\bibinfo{author}{\bibfnamefont{S.}~\bibnamefont{De~Franceschi}},
  \bibinfo{author}{\bibfnamefont{S.}~\bibnamefont{Sasaki}},
  \bibinfo{author}{\bibfnamefont{J.~M.} \bibnamefont{Elzerman}},
  \bibinfo{author}{\bibfnamefont{W.~G.} \bibnamefont{Van-Der-Wiel}},
  \bibinfo{author}{\bibfnamefont{S.}~\bibnamefont{Tarucha}}, \bibnamefont{and}
  \bibinfo{author}{\bibfnamefont{L.~P.} \bibnamefont{Kouwenhoven}},
  \bibinfo{journal}{Phys. Rev. Lett.} \textbf{\bibinfo{volume}{86}},
  \bibinfo{pages}{878} (\bibinfo{year}{2001}).

\bibitem[{\citenamefont{Yamada et~al.}(2003)\citenamefont{Yamada, Stopa,
  Hatano, Ota, Yamaguchi, and Tarucha}}]{Yamada2003}
\bibinfo{author}{\bibfnamefont{K.}~\bibnamefont{Yamada}},
  \bibinfo{author}{\bibfnamefont{M.}~\bibnamefont{Stopa}},
  \bibinfo{author}{\bibfnamefont{T.}~\bibnamefont{Hatano}},
  \bibinfo{author}{\bibfnamefont{T.}~\bibnamefont{Ota}},
  \bibinfo{author}{\bibfnamefont{T.}~\bibnamefont{Yamaguchi}},
  \bibnamefont{and} \bibinfo{author}{\bibfnamefont{S.}~\bibnamefont{Tarucha}},
  \bibinfo{journal}{Superlatt. Microstruct.} \textbf{\bibinfo{volume}{34}},
  \bibinfo{pages}{185} (\bibinfo{year}{2003}).

\bibitem[{\citenamefont{Nygard et~al.}(2000)\citenamefont{Nygard, Cobden, and
  Lindelof}}]{Nygard2000}
\bibinfo{author}{\bibfnamefont{J.}~\bibnamefont{Nygard}},
  \bibinfo{author}{\bibfnamefont{D.~H.} \bibnamefont{Cobden}},
  \bibnamefont{and} \bibinfo{author}{\bibfnamefont{P.~E.}
  \bibnamefont{Lindelof}}, \bibinfo{journal}{Nature}
  \textbf{\bibinfo{volume}{408}}, \bibinfo{pages}{342} (\bibinfo{year}{2000}).

\bibitem[{\citenamefont{Buitelaar et~al.}(2002)\citenamefont{Buitelaar,
  Bachtold, Nussbaumer, Iqbal, and Sch\"onenberger}}]{Buitelaar2002}
\bibinfo{author}{\bibfnamefont{M.~R.} \bibnamefont{Buitelaar}},
  \bibinfo{author}{\bibfnamefont{A.}~\bibnamefont{Bachtold}},
  \bibinfo{author}{\bibfnamefont{T.}~\bibnamefont{Nussbaumer}},
  \bibinfo{author}{\bibfnamefont{M.}~\bibnamefont{Iqbal}}, \bibnamefont{and}
  \bibinfo{author}{\bibfnamefont{C.}~\bibnamefont{Sch\"onenberger}},
  \bibinfo{journal}{Phys. Rev. Lett.} \textbf{\bibinfo{volume}{88}},
  \bibinfo{pages}{156801} (\bibinfo{year}{2002}).

\bibitem[{\citenamefont{Held et~al.}(1999)\citenamefont{Held, L\"uscher,
  Heinzel, Ensslin, and Wegscheider}}]{Held1999}
\bibinfo{author}{\bibfnamefont{R.}~\bibnamefont{Held}},
  \bibinfo{author}{\bibfnamefont{S.}~\bibnamefont{L\"uscher}},
  \bibinfo{author}{\bibfnamefont{T.}~\bibnamefont{Heinzel}},
  \bibinfo{author}{\bibfnamefont{K.}~\bibnamefont{Ensslin}}, \bibnamefont{and}
  \bibinfo{author}{\bibfnamefont{W.}~\bibnamefont{Wegscheider}},
  \bibinfo{journal}{Appl. Phys. Lett.} \textbf{\bibinfo{volume}{75}},
  \bibinfo{pages}{1134} (\bibinfo{year}{1999}).

\bibitem[{\citenamefont{L\"uscher et~al.}(1999)\citenamefont{L\"uscher, Fuhrer,
  Held, Heinzel, Ensslin, and Wegscheider}}]{Luscher1999}
\bibinfo{author}{\bibfnamefont{S.}~\bibnamefont{L\"uscher}},
  \bibinfo{author}{\bibfnamefont{A.}~\bibnamefont{Fuhrer}},
  \bibinfo{author}{\bibfnamefont{R.}~\bibnamefont{Held}},
  \bibinfo{author}{\bibfnamefont{T.}~\bibnamefont{Heinzel}},
  \bibinfo{author}{\bibfnamefont{K.}~\bibnamefont{Ensslin}}, \bibnamefont{and}
  \bibinfo{author}{\bibfnamefont{W.}~\bibnamefont{Wegscheider}},
  \bibinfo{journal}{Appl. Phys. Lett.} \textbf{\bibinfo{volume}{75}},
  \bibinfo{pages}{2452} (\bibinfo{year}{1999}).

\bibitem[{\citenamefont{Schoeller and
  Sch\"on}(1994{\natexlab{a}})}]{Schoeller94:18436}
\bibinfo{author}{\bibfnamefont{H.}~\bibnamefont{Schoeller}} \bibnamefont{and}
  \bibinfo{author}{\bibfnamefont{G.}~\bibnamefont{Sch\"on}},
  \bibinfo{journal}{Phys. Rev. B} \textbf{\bibinfo{volume}{50}},
  \bibinfo{pages}{18436} (\bibinfo{year}{1994}{\natexlab{a}}).

\bibitem[{\citenamefont{Tews}(2004)}]{Tews04:xx}
\bibinfo{author}{\bibfnamefont{M.}~\bibnamefont{Tews}}, \bibinfo{journal}{Ann.
  Phys.} \textbf{\bibinfo{volume}{13}}, \bibinfo{pages}{249}
  (\bibinfo{year}{2004}).

\bibitem[{\citenamefont{Cohen et~al.}(1962)\citenamefont{Cohen, Falicov, and
  Phillips}}]{cohen:04:1962}
\bibinfo{author}{\bibfnamefont{M.~H.} \bibnamefont{Cohen}},
  \bibinfo{author}{\bibfnamefont{L.~M.} \bibnamefont{Falicov}},
  \bibnamefont{and} \bibinfo{author}{\bibfnamefont{J.~C.}~\bibnamefont{Phillips}},
  \bibinfo{journal}{Phys. Rev. Lett.} \textbf{\bibinfo{volume}{6}},
  \bibinfo{pages}{316} (\bibinfo{year}{1962}).

\bibitem[{\citenamefont{Jauho}(1998)}]{jauho:1998}
\bibinfo{author}{\bibfnamefont{A.~P.} \bibnamefont{Jauho}},
  \emph{\bibinfo{title}{in Theory of Transport Properties of Semiconductor
  Nanostructures, edited by Eckehard Sch\"oll}} (\bibinfo{publisher}{Chapman
  and Hall}, \bibinfo{address}{London}, \bibinfo{year}{1998}).

\bibitem[{\citenamefont{Funabashi et~al.}(1999)\citenamefont{Funabashi,
  Ohtsubo, Eto, and Kawamura}}]{funabashi:01:1999}
\bibinfo{author}{\bibfnamefont{Y.}~\bibnamefont{Funabashi}},
  \bibinfo{author}{\bibfnamefont{K.}~\bibnamefont{Ohtsubo}},
  \bibinfo{author}{\bibfnamefont{M.}~\bibnamefont{Eto}}, \bibnamefont{and}
  \bibinfo{author}{\bibfnamefont{K.}~\bibnamefont{Kawamura}},
  \bibinfo{journal}{Jpn. J. Appl. Phys.} \textbf{\bibinfo{volume}{38}},
  \bibinfo{pages}{388} (\bibinfo{year}{1999}).

\end{thebibliography}

\end{document}